# WHEN HOLOGRAPHY MEETS COHERENT DIFFRACTION IMAGING

**Tatiana Latychevskaia,**[*] **Jean-Nicolas Longchamp, and Hans-Werner Fink**

*Institute of Physics, University of Zurich, Winterthurerstrasse 190, CH-8057, Switzerland*

[*]*tatiana@physik.uzh.ch*

## ABSTRACT

The phase problem is inherent to crystallographic, astronomical and optical imaging where only the intensity of the scattered signal is detected and the phase information is lost and must somehow be recovered to reconstruct the object's structure. Modern imaging techniques at the molecular scale rely on utilizing novel coherent light sources like X-ray free electron lasers for the ultimate goal of visualizing such objects as individual biomolecules rather than crystals. Here, unlike in the case of crystals where structures can be solved by model building and phase refinement, the phase distribution of the wave scattered by an individual molecule must directly be recovered. There are two well-known solutions to the phase problem: holography and coherent diffraction imaging (CDI). Both techniques have their pros and cons. In holography, the reconstruction of the scattered complex-valued object wave is directly provided by a well-defined reference wave that must cover the entire detector area which often is an experimental challenge. CDI provides the highest possible, only wavelength limited, resolution, but the phase recovery is an iterative process which requires some pre-defined information about the object and whose outcome is not always uniquely-defined. Moreover, the diffraction patterns must be recorded under oversampling conditions, a pre-requisite to be able to solve the phase problem. Here, we report how holography and CDI can be merged into one superior technique: holographic coherent diffraction imaging (HCDI). An inline hologram can be recorded by employing a modified CDI experimental scheme. We demonstrate that the amplitude of the Fourier transform of an inline hologram is related to the complex-valued visibility, thus providing information on both, the amplitude and the phase of the scattered wave in the plane of the diffraction pattern. With the phase information available, the condition of oversampling the diffraction patterns can be relaxed, and the phase problem can be solved in a fast and unambiguous manner. We demonstrate the reconstruction of various diffraction patterns of objects recorded with visible light as well as with low-energy electrons. Although we have demonstrated our HCDI method using laser light and low-energy electrons, it can also be applied to any other coherent radiation such as X-rays or high-energy electrons.

## 1. INTRODUCTION

The so-called phase problem is inherent to crystallographic, astronomical and optical imaging, or more general, to all scattering experiments independent of the radiation used. The distribution of the scattered wave is complex-valued with the phase part carrying information about individual scattering events, hence about the positions of elements constituting the object. The predicament is that the detector only records the intensity, i.e. the square of the amplitude of the scattered wave. Since no phase sensitive detector exists, the phase information is lost and must somehow be recovered to reconstruct the object's anatomy.

As of today, two well-studied solutions to the phase problem are prominent, in chronological order: holography[1-2] and coherent diffraction imaging (CDI)[3]. Holography was invented by Dennis Gabor in 1947[1-2]; in his original experimental scheme, an object is placed into a divergent wave (reference wave) and part of the wave is scattered by the object (object wave). No lenses are employed to focus the image at the detector plane, but instead, both waves propagate freely behind the object and their interference pattern is recorded at the detector and called the hologram. The reconstruction of the scattered complex-valued object wave is straight forward and is provided by a well-defined reference wave. In CDI, on the other hand, a plane wave impinges onto an object and only the scattered wave is detected in the far-field. The experimental arrangement is similar to crystallographic experiments, only instead of a crystal an individual molecule is placed into beam. As in holography, no lenses are employed between the sample and the detector in CDI which rules out aberrations. Yet, the result is not a direct image of the object but its diffraction pattern (squared amplitude of the scattered wave in the far-field) which is formed in the detector plane. For that reason, CDI requires the retrieval of the missing phase distribution, which is conventionally achieved by an iterative reconstruction procedure provided some prior information about the object is available[4]. The achieved resolution is the highest possible resolution[5] that can theoretically approach that obtained in crystallographic experiments. The envisioned ultimate application of CDI is to obtain an image of an individual biological molecule at atomic resolution. Some biological specimens have already been imaged by CDI employing coherent X-rays[6-13]. The first results from the Linac Coherent Light Source X-FEL facility were reported and demonstrated imaging an individual unstained mimivirus (800 nm in diameter) with 6.9 Å wavelength X-rays at 32 nm resolution[14]. Overviews of the CDI applications are presented in works[15-16].

Here, we show how holography and CDI can be merged into one superior technique: holographic coherent diffraction imaging (HCDI), which inherits the benefits of both techniques, namely the straightforward unambiguous recovery of the phase distribution and the visualization of a non-crystalline object at the highest possible resolution.

## 2. PHASE RECOVERY BY CDI

In CDI, a plane wave is scattered by an object and the wavefront distribution at a plane right behind an object is described as the exit wave $t(x,y)$. The distribution of the scattered wave in the far-field is given by the Fourier-transform of the exit wave and represents complex-valued visibility $V(X,Y) = \text{FT}(t(x,y))$. Here, $(x, y)$ and $(X, Y)$ are the coordinates in the object and detector plane respectively. The measured intensity, or diffraction pattern, is given by $I(X,Y) = |V(X,Y)|^2$. In 1952 Sayre[17] noted that the structure, whose Fourier spectrum corresponds to $|\text{FT}(t(x,y))|^2$ is the Patterson map of a single unit; this structure also has a size corresponding to twice the width of the unit cell. Therefore, if $|\text{FT}(t(x,y))|^2$ is sampled at twice the Nyquist frequency (later referred to as "oversampling"), the recovery of the structure of a crystal unit cell from its X-ray diffraction pattern alone is in principle possible. Yet, in Sayre's own words "the paper did not, however, suggest an effective way of obtaining such sampling"[18], and neither did it suggest an algorithm for the reconstruction of such images. The first algorithm of phase recovery from diffraction patterns came from electron microscope imaging. In 1972, Gerchberg and Saxton[19] proposed a practical algorithm for recovery of the complex-valued wavefront from two intensity measurements (that are feasible to obtain in a TEM) in the object and diffraction plane. Although it was the first practical algorithm for phase recovery, the Gerchberg and Saxton (GS) algorithm in its original form has been often criticized as "not entirely satisfactory, since neither its convergence nor its uniqueness is guaranteed a priori"[20] and subsequently, many different improved variations of the GS algorithm were proposed. One of the most influential improvement of the GS algorithm was proposed in 1978 by Fienup[21], who addressed the more common experimental situation when only a diffraction pattern is available and some additional information about the object is at hand, such as: the object is real, positive and localized. Most of the modern phase retrieval algorithms are based on Fienup's iterative routines, but there is still an ongoing search for faster converging algorithms providing an unambiguously defined solution[22-23].

## 3. UNIQUENESS OF THE SOLUTION

One of the important issues in CDI is the uniqueness of the solution. In 1964, it was demonstrated by Hofstetter[24] that there is no unique solution to the one-dimensional phase retrieval problem. In 1976, Huiser et al.[25-26] studied the uniqueness of solutions in the two-dimensional phase problem obtained by using the GS algorithm, concluding that a unique solution can be retrieved. After Fienup's algorithms were reported in 1978[21], the issue of finding a unique solution of a two-dimensional real and positive distribution had been studied by Bruck and Sodin[27], as well as by Bates et al.[28-29]. It was found that a unique solution can be found provided the diffraction pattern is oversampled at twice the Nyquist frequency in each dimension at the least[29-30]. However, the presence of noise in experimental diffraction patterns often leads to non-unique solutions or to a stagnation of the iterative reconstruction process at some partly reconstructed object structure[31]. In practice, results of hundreds of iterative runs are required; the final reconstruction is obtained by either averaging over all results or by subjectively choosing one and defining it as the correct reconstruction[32-34].

## 4. COMPLEX-VALUED OBJECTS

Researchers who first addressed the phase problem were mainly concerned with crystallographic and astronomical applications. To that effect, imposing the constraint that the object be real and positive seemed logical. However, CDI was also later applied for imaging specimens whose interaction with the radiation could only be described by a complex-valued transmission function (as for instance, when a cell is imaged by soft X-rays[7]). The phase problem then involved complex-valued objects as well, hence engaging a wider range of scientists in the quest for phase retrieval, not only for its being a mathematical predicament but also a concrete problem[35-36]. Nowadays, it has become common knowledge that complex-valued objects can be reconstructed when the support is accurately known, and the support boundaries are as tight as possible[37-38].

## 5. EXPERIMENTAL ISSUES IN CDI

There are several other problems associated with the CDI method. The phase retrieval is only possible provided the diffraction pattern is oversampled. The geometry of the experimental setup must thus be designed in such a way as to fulfill the oversampling condition. Furthermore, the oversampling condition in the detector plane corresponds to zero-padding in the object plane which requires the sample to be surrounded by a support with known transmission properties. For instance, when imaging a biological molecule, it must ideally be either levitating or resting on a homogeneous transparent film such as graphene[39-40].

Another problem is the missing signal in the central overexposed region of the diffraction pattern. The intensity ratio between the central spot and the signal at the rim of the detector can reach values of $10^7$; commonly used 16 bit cameras are simply not capable to capture the whole intensity range and the central (low-resolution) part is usually sacrificed by being blocked. This missing data are usually obtained by recording a low-resolution image by some other technique, a TEM image for example [4], or by recording a set of images at different exposure times[7,33].

## 6. MODIFIED CDI TECHNIQUES, FRESNEL CDI

An intensive search for better techniques has already initiated a number of novel experimental designs, such as Fresnel coherent diffraction imaging[41-44], Fourier transform holography[45-47] and ptychographic coherent diffraction imaging[48-52]. Here we would like to especially mention Fresnel or curved beam coherent diffraction imaging (FCDI)[41-44], since it uses an embedded holographic part. FCDI employs a slightly divergent (1-2°) beam, which leads to the presence of an inline hologram in the center of the diffraction pattern. The available holographic information immediately solves such problems as finding a mask for the object support for iterative phase retrieval; it provides fast convergence of the iterative reconstruction and uniqueness of the found solution. However, there are several drawbacks of this method. The average intensity in the central "holographic" region is about $10^4$ times higher than the intensity of the "diffraction" part of the pattern. To record both parts in one image one would need a set of images under different exposures times (including a very long exposure to record the signal at the rim of the detector). However, since here the diffraction pattern is not simply the Fourier transform of the object function (the incident wavefront is not planar but spherical), it is highly sensitive to any lateral shifts. Thus, a longer acquisition time intended to increase the contrast of higher order diffraction signals in the diffraction pattern, in fact, just blurs out the high order diffraction information. Another possibility to properly record both, the holographic and the diffraction part of the pattern in one exposure, is to use a 20 bit or higher dynamic range detector. Finally, using a detector of fixed size the curvature of the incident wave leads to decreased resolution in comparison to conventional CDI.

## 7. RELATION BETWEEN INLINE HOLOGRAPHY AND CDI

The inline holographic scheme can relatively easily be converted to the CDI experimental scheme by modifying the incident wavefront from divergent to parallel. The reconstruction of an object distribution $t(x,y)$ from its inline hologram $H(X,Y)$ requires a deconvolution with the impulse response of a free space propagation factor:

$$S^*(u,v)\text{FT}(H(X,Y)) \approx \text{FT}(t(x,y)), \qquad (1)$$

where $S(u,v)$ is the Fourier transform of the free space propagation factor (see Appendix A). The right side of Eq. 1 is the visibility $V(X,Y) = \text{FT}(t(x,y))$ - the distribution of the scattered wave in the far-field introduced earlier.

In CDI, the measured intensity in the far-field is given by:

$$I(x,y) = |V(X,Y)|^2 = |\text{FT}(t(x,y))|^2. \qquad (2)$$

Equations 1 and 2 demonstrate that the modulus of the Fourier transform of the hologram $|\text{FT}(H(X,Y))|$ (which we will also call Fourier spectrum of the hologram) delivers the amplitude of the scattered object wave in the far-field $|\text{FT}(t(x,y))|$. In the following we investigate practical consequences of this concept.

We performed a set of optical experiments verifying the notion that the amplitude of the Fourier spectrum of a hologram corresponds indeed to the amplitude of the diffraction pattern. The experimental optical scheme which allows an easy alteration between diffraction pattern and hologram acquisition modes is shown in Fig. 1. As a test sample we used two tungsten wires of 15 μm in diameter twisted around each other, see Fig. 2(a). The shape of the sample was chosen to mimic the DNA double helical structure and thus to obtain an optical diffraction pattern resembling the famous X-ray fiber diffraction pattern of DNA obtained by Rosalind Franklin. A hologram and a diffraction pattern of the sample were recorded and are shown in Fig. 2(b) and (c). The digital Fourier transform of the hologram is shown in Fig. 2(d). The Fourier spectrum of the hologram lacks the higher orders due to the presence of noise (see Appendix B). The resolution obtained in the hologram can be directly estimated by inspecting the highest detectable frequencies in the Fourier spectrum of the hologram and it corresponds to 6 μm (indicated in the blue circle). It is evident that the Fourier spectrum of the hologram resembles the measured diffraction pattern shown in Fig. 2(c). Here, the diffraction pattern provides the same resolution as the hologram - namely 6 μm, but it is recorded while fulfilling the oversampling condition. Therefore, one can conclude that if the phase distribution was readily available, thus eliminating the need to oversample diffraction pattern, diffraction pattern exhibiting even higher modes could be recorded using the screen of the same size.

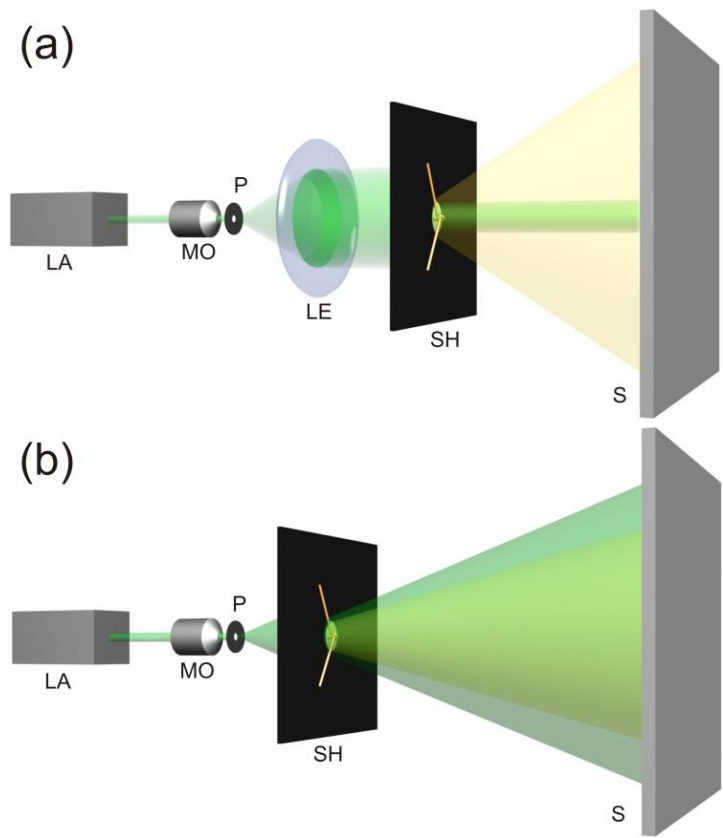

*Fig. 1. Optical schemes. (a) Scheme for recording diffraction patterns. Laser light (LA) passes the spatial filtering system (consisting of a microscope objective (MO) with a numerical aperture $NA = 0.85$ and a pinhole (P) of diameter 10 μm). The object is fixed at the sample holder (SH). For the recording of diffraction patterns, the wavefront is broadened by a lens (LE). (b) Scheme for recording holograms. The lens LE is removed and the sample is shifted closer to the pinhole P. Both, diffraction patterns and holograms are recorded on screen (S) using a 10 bit CCD camera with 1000x1000 pixel$^2$.*

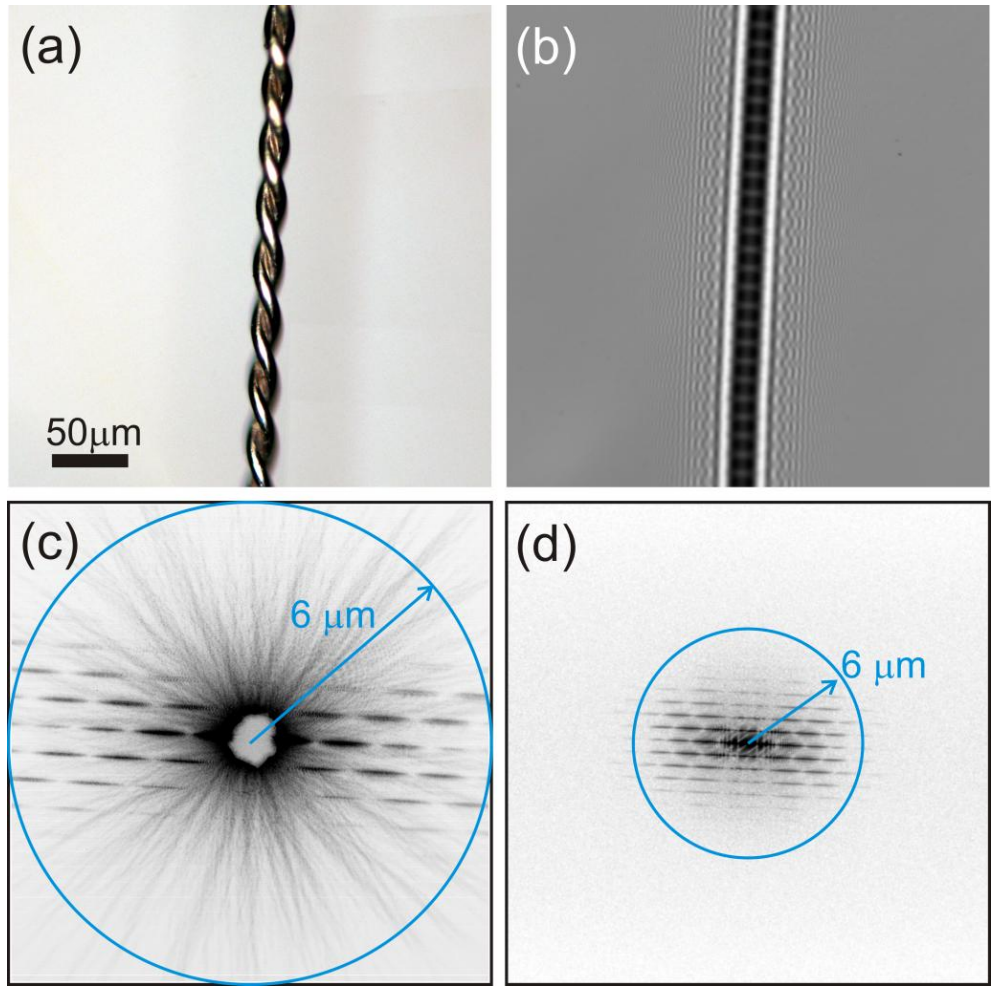

*Fig. 2. Experimental verification of the relation between a hologram and a diffraction pattern. (a) Reflected-light microscopy image of two twisted tungsten wires. (b) Hologram recorded using the scheme shown in Fig. 1(b) with a pinhole-screen distance of 100 mm and a pinhole-sample distance of 5.8 mm. (c) Diffraction pattern recorded using the scheme shown in Fig. 1(a) at a sample-screen distance of 145 mm. Both patterns were recorded on a 25x25 mm$^2$ screen and sampled with 1000x1000 pixel$^2$. (d) The amplitude of the Fourier transform of the hologram is displayed using a logarithmic and inverted intensity scale.*

## 8. HCDI

To set the stage for our HCDI method, we like to begin by quoting Gerchberg and Saxton[19] who, in 1972, made the following comment on using a random phase as the initial guess in the iterative phase retrieval: "This is not necessary in every case and indeed there is every reason to suppose that an educated guess at the correct phase distribution would lessen the computation time required for the process to achieve an acceptable squared error." That was also confirmed by Bates[53] in 1978: "It is seen that the availability of even inaccurate phase data can greatly simplify the computational procedure…"

Accurate phase information is indeed exactly what is provided by holography. Thus, in HCDI two images are recorded: a diffraction pattern and an inline hologram of the same object. The utilization of these two images must not be confused with other methods which also require two images, as in the GS algorithm, where a diffraction pattern and a low-resolution image of the object are used. Since the low-resolution image of the object is always a real-valued image, the phase estimate obtained from its Fourier transform will always stagnate the reconstruction process of some real-valued object. Hence, especially for objects with complex-valued transmission functions, a low-resolution image of the object cannot be used for guessing the phase distribution in the detector plane. A hologram, in turn, intrinsically contains information on both phase distributions, that in the detector plane and that in the object plane[54], and therefore provides a quite accurate phase estimation for starting the iterative phase retrieval process.

## 8.1 THE CENTRAL SPOT

Since the Fourier transform of the hologram is the amplitude of the visibility, this feature immediately solves the problem of the commonly missing information in the central spot of the diffraction pattern. It can now be directly obtained from the Fourier spectrum of the hologram $|\text{FT}(H(X,Y))|$.

## 8.2 THE OVERSAMPLING CONDITION

Here, we would like to recall that Sayre[17] and Bates et al.[28-29] arrived at the "oversampling" condition following the fact that the Fourier transform of the measured intensity distribution is the autocorrelation of the object distribution. Since the autocorrelation function is always at least twice the size of the object itself, to reconstruct the autocorrelation unambiguously, the intensity of the diffraction pattern must be oversampled by a factor of 2. In our method, the phase of $\text{FT}(H(X,Y))S^*(u,v)$ provides an accurate estimate of the phase distribution for the iterative reconstruction. Therefore, the oversampling condition becomes obsolete. This can be illustrated by counting the number of unknowns and available equations. The number of unknowns for the real and imaginary parts of the object wave amounts to $2N^2$. The number of equations provided by a holographic measurement is $N^2$ and that provided by detecting a diffraction pattern is also given by $N^2$. Therefore, a hologram and a diffraction pattern combined provide a matching number of equations and unknowns, and thus ensure a unique solution. The elimination of the oversampling condition leads to a certain freedom of choice of experimental parameters, which can now be optimized for achieving the best resolution. Of course, the digital recording of a diffraction pattern must still satisfy the condition of sampling at least at the Nyquist frequency in the object domain. Next, we show how this available phase information can be fed into the reconstruction routine.

## 9. HCDI. SIMULATED EXAMPLES

As a test sample, some text of a piece of literature was selected. This allows judging the resolution in the reconstructed images by visual inspection. The chosen text is a part of the Mark Twain novel "A Fable". The text placed inside a circle imitating an aperture and was sampled with 1000x1000 pixels, as shown in Fig. 3(a). The hologram as well as the diffraction pattern was simulated using realistic parameters that are later also applied in an experimental study. The hologram parameters are: wavelength - 532 nm, sample size - 1.6x1.6 mm$^2$, source-sample distance - 5.33 mm, source-screen distance - 1 m, and screen size - 300 mm. The diffraction pattern was calculated using the digitized form of the Huygens-Fresnel diffraction integral in the Fraunhofer approximation:

$$I(p,q) = \left| \sum_{n,m}^{N} t(n,m) \exp\left(-\frac{2\pi i}{\lambda z}(mp+nq)\right) \right|^2, \quad (3)$$

where $\lambda$ is the wavelength, $z$ the distance between sample and screen and $t(n,m)$ the digitized transmission function of the object (as shown in Fig. 3(a)) which was zero-padded to 2000x2000 pixels$^2$, so that n,m=1…2000. The number of pixels in the detector plane was selected to be 1000x1000 pixels$^2$, so that p,q=1…1000. Instead of using a fast Fourier transformation, the intensity at each pixel in the detector plane (p,q) was obtained by calculating the squared amplitude of the sum of the scattered waves over all pixels in the object domain as provided by Eq. 3. Note that the number of pixels in the object and screen plane are different. The parameters of the calculated diffraction pattern are: sample size - 3.2x3.2 mm$^2$, sample-screen distance - 905 mm, and screen size - 300 mm. Both, the hologram and the diffraction pattern were sampled with 1000x1000 pixel$^2$.

An obvious way to implement the available phase distribution from the hologram is to use it for initializing the Fienup iterative routine instead of starting with a random phase distribution as conventionally done. For the first iteration, the complex-valued visibility is composed as follows: the amplitude is given by the square root of the measured intensity; the phase, according to Eq. 1, is given by the phase of $FT(H)S^*(u,v)$; we refer to Appendix A for a detailed explanation of $S^*(u,v)$. It amounts to $\exp\left((i\pi\lambda D^2)(n^2+m^2)/dS^2\right)$, where D is the distance between the origin of the spherical wave and the screen, d is the distance between the source and the object, S is the screen size, and (m,n) denote the pixel position. In the iterative process, the approximation given by Eq. 1 turns into an equation, and the artifact terms, such as the zero-order and the twin image terms, eventually fade away.

To study the effect of replacing the central part of the diffraction pattern with that of the Fourier spectrum of the hologram, the central region of 22 pixels diameter of the square root of the diffraction pattern was replaced by the isomorphic central region of the Fourier spectrum of the hologram for the first 100 iterations (we note that when it is replaced in just the first iteration or in all iterations, the final result is almost the same). For the remaining iterations, the central region was not subject to any constraints. In the object domain we applied the constraint of non-negative absorption [55-56] while the phase distribution remained unconstrained. These conditions allow for recovery of objects exhibiting a complex-valued transmission function.

The results of the reconstruction after 300 iterations are shown in Fig. 3(b). Without any support mask (Fig. 3(b), left) the object text is almost perfectly recovered, though some artefacts remain. By employing a loose support mask of about 8 pixels away from the object contour (Fig. 3(b), right), the object is recovered almost free of any artifact. The error function, displayed in Fig. 3(c), was calculated as:

$$\text{Error} = \frac{\sum \left||U_0| - |U_i|\right|}{\sum |U_0|}, \qquad (4)$$

where $|U_0|$ is the square root of the measured intensity and $|U_i|$ is the updated amplitude of the complex wave at the detector plane after each iteration. The error function decreases rapidly with the number of iteration followed by an asymptotic approach towards zero, as displayed in Fig. 3(c). After 300 iterations, the error reaches a value of 0.007 with a support mask applied, and 0.003 without a support mask.

We have hereby demonstrated a way to recover a non-oversampled diffraction pattern provided an inline hologram of the sample is available in addition to the diffraction pattern.

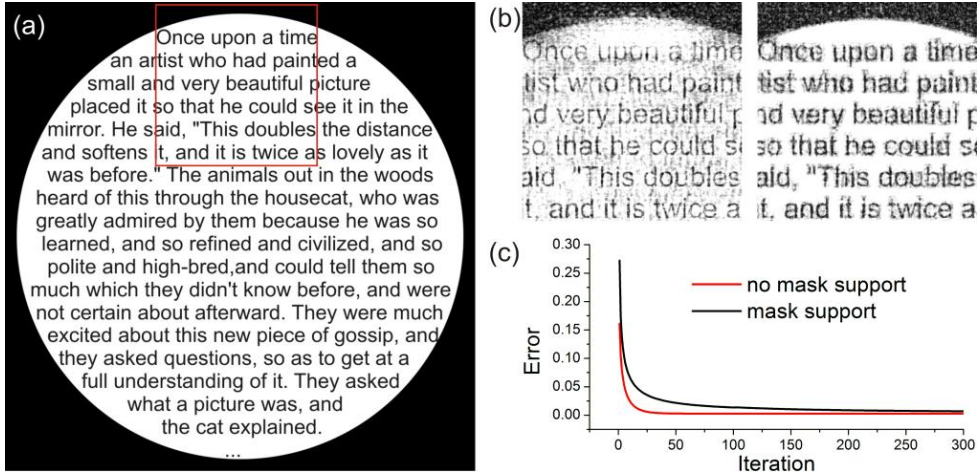

*Fig. 3. Demonstration of the HCDI method using a simulation. (a) The transmission function of the sample, sampled with 1000x1000 pixels. The red square marks the part of the sample for which the reconstruction is shown below. (b) Reconstruction obtained after 300 iterations. Left: No support mask was employed. Right: A support mask was used for all iterations. (c) Error as a function of iteration.*

## 10. HCDI. Experimental examples

### 10.1 HCDI. Optical experiments

An optical setup employing laser light of 532 nm wavelength has been used for verifying the aforementioned methods. It allows recording a hologram and a diffraction pattern of the same object without significant modification, as shown in Fig. 1: only the illuminating wavefront is converted from parallel to spherical. Two objects were selected: (i) a knotted hair because of its simple geometry and easy-to-interpret diffraction pattern, and (ii) a wing of a fly, which results in a diffraction pattern without distinct peaks. The samples are placed over holes in a material with zero transmission. This provides the effect of an apodization function and helps avoiding artefacts at the rim of the reconstruction. The sample sizes were 1.6x1.6 mm$^2$, the screen size was 300x300 mm$^2$. For recording holograms, the source-sample distance was 5.3 mm and the source-screen distance was 1 m. For recording the diffraction patterns the sample-screen distance was 905 mm. Both, holograms and diffraction patterns were sampled with 1000x1000 pixel$^2$.

Figure 4 shows the hologram and the diffraction pattern of the knotted hair (Fig. 4(b) and (c)). The squared amplitude of the Fourier transform of the hologram, displayed in Fig. 4(e), exhibits a distribution which compares well to the measured diffraction pattern shown in Fig. 4(d). The low-order modulations in $|\text{FT}(H(X,Y))|$ and in the diffraction pattern appear at the same positions. The higher order diffraction spots are absent in $|\text{FT}(H(X,Y))|$ owing to experimental noise in the reference wave. A hologram can be reconstructed on its own, and the results of such reconstructions are displayed in Fig. 4(e) and (f). The details in these hologram reconstructions are blurred and the presence of the out-of-focus twin image is apparent.

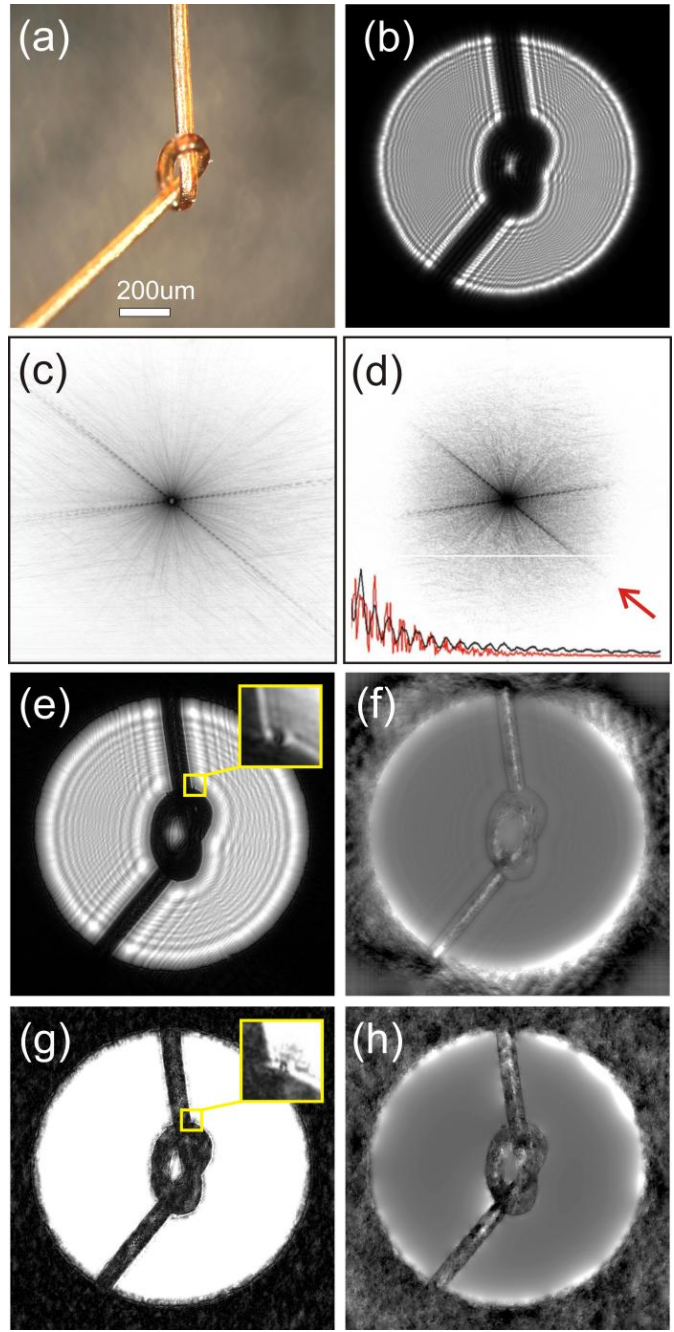

*Fig. 4. HCDI reconstructions of a human hair. (a) Reflected-light microscopy image of a hair. (b) Recorded hologram: the pinhole-sample distance amounts to 5.3 mm. (c) Experimental diffraction pattern recorded at 1 m from the sample. (d) Amplitude of the Fourier transform of the hologram. The inset shows the profiles of the square root of the experimental diffraction pattern intensity (black) and the amplitude of $\mathrm{FT}(H)$ (red) along the direction indicated by a red arrow. The resolution provided in a hologram can be estimated by examining the amplitude of $\mathrm{FT}(H)$, and it is usually less than the resolution provided in the diffraction pattern. (e) Amplitude distribution reconstructed from the hologram. The superimposed interference pattern arises from the twin image. In the inset, a magnified 40x40 pixel$^2$ part of the reconstruction is shown. (f) Reconstructed phase distribution from the hologram. (g) Reconstructed amplitude distribution using HCDI. The inset shows a magnified 40x40 pixel$^2$ part of the reconstruction, demonstrating an improved resolution in comparison to the hologram reconstruction. (h) Reconstructed phase distribution using HCDI.*

The recorded diffraction pattern was reconstructed using the phase information available from the hologram as described above. For the first loop of the iterative phase retrieval procedure the complex-valued distribution in the detector plane is composed from the measured amplitude and phase of $\text{FT}(H(X,Y))S^*(u,v)$. This phase distribution is used only for the first iteration. The missing central region of 18 pixels in diameter is filled with the amplitude distribution of $\text{FT}(H)$ for the first 100 iterations. A non-negative absorption constraint [55] and a loose support mask of about 20 pixels apart from the object contour is applied in the object plane. A recognizable object distribution is already achieved just after the first iteration. The amplitude and phase distributions retrieved after 300 iterations are shown in Fig. 4(g) and (h). Fine details of about 1 pixel in size (corresponding to 1.6 µm) can be resolved in the reconstruction displayed in Fig. 4(g).

Another application of the HCDI method is shown in Fig. 5. Here, the object, a wing of a small fly, was selected for the sake of obtaining a less structured diffraction pattern without distinct peaks spots. Again, the measured diffraction pattern and the squared amplitude of the Fourier transform of the hologram are visually identical. The amplitude and phase distributions retrieved after 300 iterations by the same HCDI iterative procedure as described above are shown in Fig. 5(e) and (f).

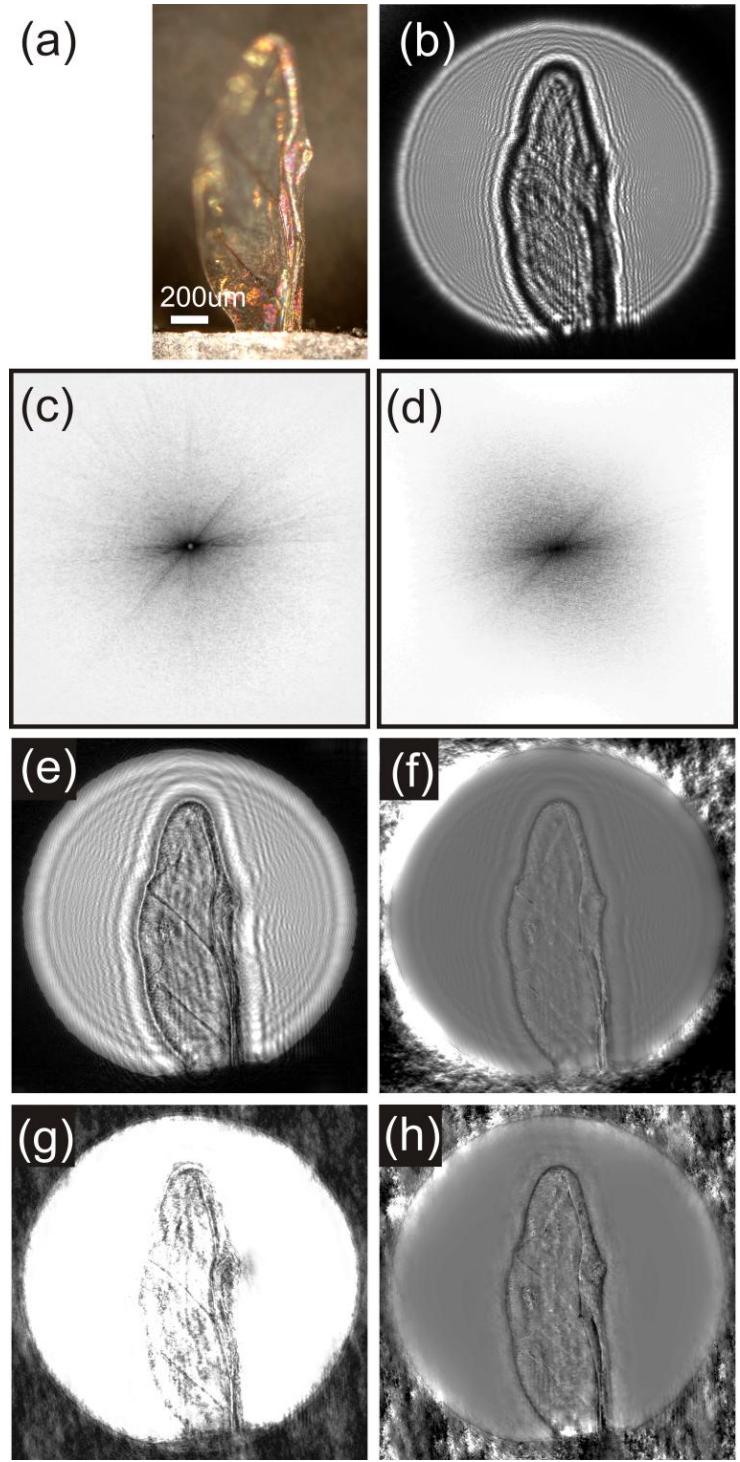

Fig. 5. HCDI reconstructions of a fly wing. (a) Reflected-light microscopy image of a small fly wing. (b) Recorded hologram: the pinhole-sample distance amounts to 5.3 mm. (c) Experimental diffraction pattern recorded at 1 m from the sample. (d) Amplitude of $\mathrm{FT}(H)$. (e) Amplitude distribution reconstructed from the hologram. (f) Phase distribution reconstructed from the hologram. (g) Reconstructed amplitude distribution using HCDI. (h) Reconstructed phase distribution using HCDI.

## 10.2 LOW-ENERGY ELECTRON EXPERIMENTS

We also tested the HCDI method on patterns recorded with coherent low-energy electrons. As a sample we used multi-walled carbon nanotubes stretched over holes milled in a thin carbon film, prepared in a manner described elsewhere[57]. The experimental setup is sketched in Fig. 6(a); the transition from a spherical to a plane wavefront was realized by adjusting the voltage at the micro-lens[57-58]. The electron hologram was recorded with electrons of 51 eV energy and is shown in Fig. 6(b); the distance between electron source and the sample was 640 nm. The diffraction pattern is composed of 4 images recorded with electrons of 145 eV energy at exposure times of 4, 8, 16 and 32 seconds, shown in Fig. 6(c). The missing central part of the diffraction pattern is filled with $\mathrm{FT}(H)$. Both, the hologram and the diffraction pattern, are sampled with 1200x1200 pixel$^2$. The images were reconstructed using 300 iterations as described above in the optical experiment section; the central region of 60 pixels diameter was replaced with the amplitude distribution of $\mathrm{FT}(H)$ for the first 100 iterations. The resulting reconstruction, shown in Fig. 6(d) is in a good agreement with the TEM image of the same sample displayed in Fig. 6(e).

## 11. CONCLUSION

We have demonstrated that an inline hologram can be recorded in a modified CDI experimental scheme. The amplitude of the Fourier transform of the inline hologram is isomorphic to the amplitude of the diffraction pattern and with this, its phase distribution provides accurate starting values for the subsequent phase retrieval routine. Given such accurate phase values to start with, the condition of oversampling diffraction patterns becomes obsolete. Experimental parameters can no longer be selected to fulfil the oversampling condition, but can instead be selected to achieve highest possible resolution. We suggested procedure for implementing such introductory phase information into the iterative phase retrieval routine by using $\mathrm{FT}(H)$ just for the first iteration. Although, we have demonstrated the HCDI method using light optics and low-energy electrons, it can clearly also be applied to other coherent radiation such as X-rays or high-energy electrons with an expected resolution scaling with the wavelength. The holographic acquisition scheme can be realized in X-ray diffraction experiments either by shifting the Fresnel zone plate or the object, thus placing the object into a divergent beam. For future experiments, as the information stored in a holographic image is actually three-dimensional, the method will potentially be extended to reconstruct the three-dimensional structure of objects from their diffraction patterns.

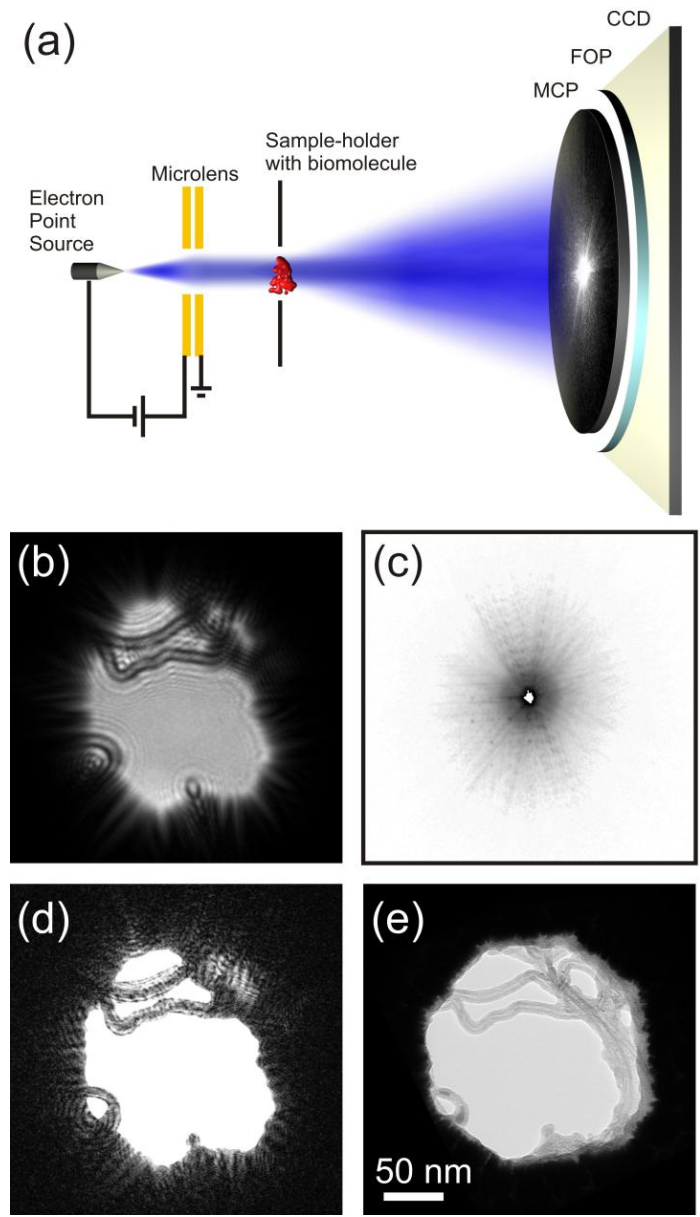

*Fig. 6. HCDI reconstructions of a coherent low-energy electron diffraction pattern of individual carbon nanotubes. (a) Schematics of the low-energy electron microscope, the distance between electron source and detector amounts to 68 mm. The detector components are: micro-channel plate (MCP), fiber optical plate (FOP) and CCD chip. (b) Hologram recorded with electrons of 51 eV kinetic energy. (c) Diffraction pattern recorded with electrons of 145 eV kinetic energy. (d) Reconstructed amplitude distribution using HCDI. (e) TEM image recorded with 80 keV electrons. In (b) and (d) the central parts of the images, with 600x600 pixel$^2$ are shown.*

## 12. APPENDIX A: INLINE HOLOGRAM RECONSTRUCTION BY DECONVOLUTION

The transmission function in the object plane can be written as:

$$t(x, y) = 1 + o(x, y) = \exp(-a(x, y))\exp(-i\phi(x, y)), \quad (A1)$$

where $a(x, y)$ and $\phi(x, y)$ describe the absorbing and phase shifting properties in some plane in the object domain. An incident spherical wave passing the object is described by $(1/z_0)\exp(i\pi/\lambda z_0(x^2 + y^2))$, where $z_0$ denotes the distance between the origin of the wave and the object. The scattered wave in the detector plane is represented by:

$$U(X,Y) = -\frac{i}{\lambda z_0 z_s} \iint t(x,y)\exp\left(\frac{i\pi}{\lambda z_0}(x^2 + y^2)\right) \times$$

$$\exp\left(\frac{i\pi}{\lambda z_s}\left((x-X)^2 + (y-Y)^2\right)\right) dxdy. \quad (A2)$$

Here, $z_s$ is the distance between the object and detector, and $(X,Y)$ denotes the coordinates in the detector plane. The intensity in the detector plane, and thus the hologram distribution, is given by:

$$H(X,Y) = |U(X,Y)|^2 = \frac{1}{(\lambda z_0 z_s)^2} \times$$

$$\left|\iint t(x,y)\exp\left(\frac{i\pi}{\lambda z_0}\left(\left(x - X\frac{z_0}{z_s}\right)^2 + \left(y - Y\frac{z_0}{z_s}\right)^2\right)\right) dxdy\right|^2. \quad (A3)$$

The expression under the integral can be represented as a convolution and leads to the following expression after hologram normalization (division by the intensity of the wave at the detector in the absence of the object)[59]:

$$H(X,Y) = |t(X,Y) \otimes s(X,Y)|^2, \quad (A4)$$

in which we introduced the Fresnel function:

$$s(x, y) = -\frac{i}{\lambda z_0}\exp\left(\frac{i\pi}{\lambda z_0}(x^2 + y^2)\right). \quad (A5)$$

Its Fourier transform $S(u,v)$ is given by:

$$S(u,v) = -\frac{i}{\lambda z_0}\iint \exp\left(\frac{i\pi}{\lambda z_0}(x^2 + y^2)\right) \times$$

$$\exp(-2\pi i(xu + yv)) dxdy = \exp(-i\pi\lambda z_0(u^2 + v^2)). \quad (A6)$$

The hologram distribution in Eq. A4 can be rewritten as:

$$H(X,Y) = \left|(1+o(X,Y)) \otimes s(X,Y)\right|^2. \tag{A7}$$

Taking into account that

$$1 \otimes s(X,Y) = -\frac{i}{\lambda z_0} \iint \exp\left(\frac{i\pi}{\lambda z_0}\left((x-X)^2 + (y-Y)^2\right)\right) dxdy = 1, \tag{A8}$$

we expand the hologram distribution as:

$$H(X,Y) = 1 + o(X,Y) \otimes s(X,Y) +$$

$$o^*(X,Y) \otimes s^*(X,Y) + \left|o(X,Y) \otimes s(X,Y)\right|^2. \tag{A9}$$

Since we know from inline holography reconstructions that the first two terms represent the dominant information on the object, for further discussion, we neglect the conjugate term and the small perturbation term for further discussions:

$$H(X,Y) \approx 1 + o(X,Y) \otimes s(X,Y). \tag{A10}$$

Reconstructing the hologram includes three steps.

1. It starts with calculating its Fourier transform:

$$\text{FT}[H(X,Y)] \approx \delta(u,v) + O(u,v)S(u,v). \tag{A11}$$

2. It is then followed by multiplying with $S^*(u,v)$ and using $|S(u,v)|^2 = 1$:

$$S^*(u,v)\text{FT}(H(X,Y)) \approx S^*(u,v)\delta(u,v) + O(u,v). \tag{A12}$$

3. Finally, a backward Fourier-transformation is applied to the result:

$$\text{FT}^{-1}\left(S^*(u,v)\text{FT}(H(X,Y))\right) = 1 + o(x,y), \tag{A13}$$

where we use:

$$\iint S^*(u,v)\delta(u,v)\exp(2\pi i(xu+yv))dxdy = 1. \tag{A14}$$

Next, we would like to point out that according to Eq. A13, the following equation applies:

$$S^*(u,v)\text{FT}(H(X,Y)) = \text{FT}(t(x,y)) \tag{A15}$$

where $\text{FT}(t(x,y))$ by definition is the visibility $V(X,Y) = \text{FT}(t(x,y))$ – the distribution of the scattered wave in the far-field.

# 13. APPENDIX B: RESOLUTION IN INLINE HOLOGRAPHY

The achievable resolution in inline holography can be illustrated by the example of a hologram of a point scatterer. On that note, an ideal noise-free inline hologram of a point-scatterer happens to be a Fresnel zone plate (FZP). Such a hologram, when recorded on an infinitely large screen, results in a δ-function in the reconstruction. However, in reality, the limited size of the screen and the finite size of the detector pixel lead to a limited resolution. The finest resolved fringes in the hologram define the resolution.

The Fourier spectrum of an inline hologram of an object, to a good approximation, is the diffraction pattern of the object, placed at a distance $z_1$ from the screen of size $S_0 = \lambda z_S N / S_H$. Here, $z_s$ is the distance between the source of the divergent wave and the screen, $S_H$ denotes the screen size and N – the number of pixels.

For large distances $z_1$, i.e. small magnification $M = z_S / z_1$, the resulting hologram is a FZP with very high NA (numerical aperture), and thus the challenge is to resolve the outermost fine fringes whose width is comparable to the detector pixel size. According to the Shannon sampling theorem, it requires at least 2 samplings per period to correctly represent a periodic signal. This implies that the finest resolved fringe must have a period of 2 pixels, independent of the size of the pixel. For any periodic signal of period $T_p$ (in pixels), peaks at $\nu_p = \pm N/T_p$ are observed in the Fourier spectrum. Thus, for $T_p = 2$ (in pixels), the corresponding peaks in the Fourier spectrum are found at $\nu_p = \pm N/2$ (in pixels). The positions of these peaks define the resolution limited by the sampling: $R_S = 2\lambda z_1 / S_0$. When substituting $S_0$ as defined above, we obtain:

$$R_S = 2 \frac{z_1 S_H}{z_S N}. \qquad (B1)$$

The resolution $R_S$ is thus linearly increasing with $z_1$.

However, the finest fringes such as those which have a period of 2 pixels only are very difficult to be distinguished from experimental noise. In practice, the Fourier spectrum of a hologram always shows a limited resolution $R_{exp} < R_S$.

As the distance $z_1$ decreases, the period of the finest fringes increases and takes up more than 2 pixels per period. The fringes are therefore correctly sampled.

The fundamental resolution limit in inline holography is given by the Abbe diffraction limit which amounts to:

$$R_{NA} = \frac{\lambda}{NA}, \qquad (B2)$$

where the numerical aperture of the setup is approximately given by $NA \approx S_H / (2 z_S)$. The resolution $R_{NA}$ is only limited by the numerical aperture of the setup and the wavelength.

To verify the experimentally obtainable resolution in inline holograms, 11 optical holograms of two twisted tungsten wires at different $z_1$ distances but same source-screen distance of $z_S = 100$ mm, and a screen size $S_H = 25$ mm, were recorded. Three of these holograms are shown in Fig. 7. The highest detected frequency in the Fourier spectrum of the hologram (shown as blue circles in Fig. 7(a)-(c)) defines the experimental resolution $R_{exp}$ which is calculated as

$$R_{\exp} = 2\frac{\lambda z_1}{S_0^*}, \tag{B3}$$

where $S_0^*$ is the "effective screen" corresponding to the area where the signal in the Fourier spectrum is still observable (the area within the blue circles). The estimated resolution $R_{\exp}$ is plotted against $z_1$ in the graph of Fig. 7(d). The resolution limited by the sampling $R_S$ is indicated by red circles in the Fourier spectra and by a red line in the graph of Fig. 7(d). The resolution given by the numerical aperture of the setup is indicated by a green line in Fig. 7(d).

The results presented in Fig. 7 lead to the following conclusions:

(i) The resolution of an inline hologram can be directly evaluated from its Fourier spectrum.

(ii) At large source-sample distances $z_1$ the resolution is limited by the sampling of the hologram.

(iii) At small source-sample distances $z_1$ the resolution is limited by the numerical aperture of the setup.

## ACKNOWLEDGEMENTS

The authors would like to acknowledge the assistance of Conrad Escher for the sample preparation. We would also like to thank Mirna Saliba for careful reading of the manuscript. The "Forschungskredit of the University of Zurich" as well as the Swiss National Science Foundation are gratefully acknowledged for their financial support.

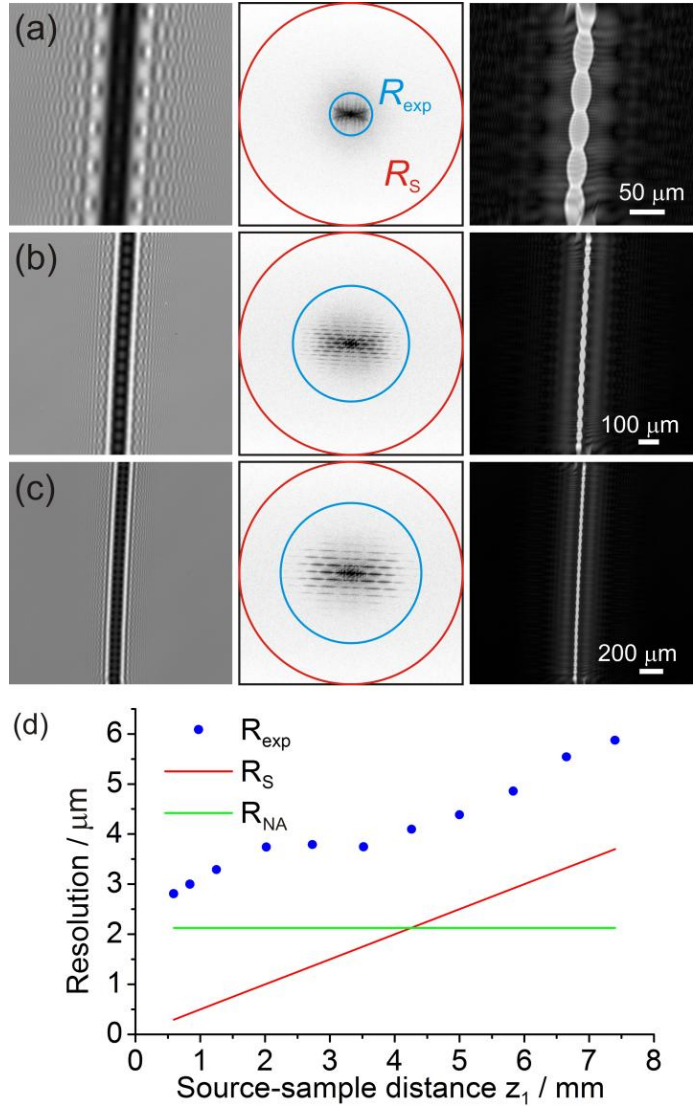

*Fig. 7. Sequence of 11 optical holograms of two twisted tungsten wires recorded with 532 nm laser light, at a source sample distance of 100 mm, a screen size of 25 mm, and different source-sample $z_1$ distances. The three holograms shown in the left column are recorded at the following source-sample distances: (a) 1.25 mm, (b) 3.50 mm and (c) 7.40 mm. In the center column, the inverted amplitudes of the Fourier spectra of the corresponding holograms (in logarithmic intensity scale) are displayed. The red circles indicate the resolution limited by the digital sampling $R_S$. The blue circles show the resolution corresponding to the highest observed frequency in the Fourier spectrum. The reconstructed objects are displayed in the right column. (d) Plot of the resolution as a function of the source-sample distance. The blue dots correspond to the experimental resolution $R_{exp}$. The red line shows the resolution limit $R_S$ given by the sampling. The green line indicates the constant resolution $R_{NA}$ limited by the numerical aperture of the setup.*